\documentclass{PoS}

\usepackage{wrapfig}

\usepackage{amsmath}
\usepackage{amsfonts,amsbsy}
\usepackage{amssymb}
\usepackage{appendix}
\usepackage{array}
\usepackage{multirow}
\def\be{\begin{equation}}
\def\ee{\end{equation}}
\def\bea{\begin{eqnarray}}
\def\eea{\end{eqnarray}}

\def\eq#1{{Eq.~(\ref{#1})}}
\def\fig#1{{Fig.~\ref{#1}}}
\newcommand{\Lb}{\left(}
\newcommand{\Rb}{\right)}

\def\beq{\begin{equation}}
\def\eeq{\end{equation}}
\def\bea{\begin{eqnarray}}
\def\eea{\end{eqnarray}}

\title{b-CGC versus IP-Sat and high precision combined HERA data}

\ShortTitle{b-CGC versus IP-Sat and high precision combined HERA data}

\author{\speaker{Amir H. Rezaeian}
\\
Departamento de F\'\i sica, Universidad T\'ecnica
Federico Santa Mar\'\i a, Avda. Espa\~na 1680,
Casilla 110-V, Valparaiso, Chile\\
and\\
Centro Cient\'\i fico Tecnol\'ogico de Valpara\'\i so (CCTVal), Universidad T\'ecnica
Federico Santa Mar\'\i a, Casilla 110-V, Valpara\'\i so, Chile\\
E-mail: \email{Amir.Rezaeian@usm.cl}}

\abstract{ The Impact-Parameter dependent Color Glass Condensate (b-CGC) and Saturation (IP-Sat) dipole models incorporate key features of small-x physics properties and match smoothly to the perturbative QCD regime at large $Q^2$ for a given $x$. Although both models include saturation effects and depend on impact-parameter, the former is based on the non-linear Balitsky-Kovchegov equation, while the latter is based on DGLAP evolution. After confronting the models to the recently released high precision combined HERA data,  we show that in both models, the typical impact-parameter probed in the total $\gamma^{*}p$ cross-section is about $b\approx 2\div 3\,\text{GeV}^{-1}$ and  the proton saturation scale is $Q_S<1$ GeV in HERA kinematics.  We show that most features of inclusive DIS and exclusive diffractive data at HERA are correctly reproduced in both models.  Nevertheless, the b-CGC and the IP-Sat models give significantly different predictions beyond the current HERA kinematics for the structure functions at very low $x$ and high virtualities $Q^2$, and for the exclusive diffractive vector meson and DVCS  production at  high $t$.  }

\FullConference{XXII. International Workshop on Deep-Inelastic Scattering and Related Subjects,\\
		28 April - 2 May 2014\\
		Warsaw, Poland}

\begin{document}

\section{Introduction}
Exclusive diffractive processes such as exclusive vector meson production or deeply virtual Compton scattering (DVCS) alongside with inclusive deep inelastic scattering (DIS) are excellent probes of the unitarity limit of QCD. An effective field theory describing the high-energy limit of QCD is the Color Glass Condensate (CGC) \cite{mv}.  A key ingredient in particle production at small-x in the CGC approach is the universal dipole amplitude, the imaginary part of the quark-antiquark scattering amplitude on a proton or nuclear target.
The choice of impact-parameter profile of the dipole amplitude entails intrinsically non-perturbative physics,  which is beyond the QCD weak-coupling approach to the CGC. It is well known that the small $x$ evolution equations generate a power law Coulomb-like tail, which is not confining at large distances \cite{al,bk-c,ana} and therefore may violate the unitarity bound. For these reasons, in practice,  supported by the $t$-distribution of the exclusive diffractive processes (for $|t|<1$), a Gaussian profile for the impact parameter dependence of the dipole amplitude is assumed.

There are two well known impact-parameter dependent dipole models in the market, the so-called b-CGC \cite{bcgc-old,bcgc-new} and IP-Sat \cite{ipsat-old,ipsat-new} models. The IP-Sat dipole amplitude  can be derived at the classical level in the CGC~\cite{mv}, contains an eikonalized gluon distribution which satisfies DGLAP evolution while explicitly maintaining unitarity. In the b-CGC dipole model,  two well-known limiting regimes are matched, the one of the BFKL equation and the region deep inside the saturation, by simple analytical interpolations \cite{IIM}.
Both models also match smoothly to the high $Q^2$ perturbative QCD limit. The b-CGC and the IP-Sat models have been both applied to various reactions, from DIS and diffractive processes \cite{bcgc-old,bcgc-new,ipsat-old,ipsat-new,ip-g1,ip-g2} to proton-proton \cite{pp-LR,pp0} and heavy ion collisions at RHIC and the LHC, see e.g. Refs.\,\cite{pA1, Schenke:2012wb}.

The main purpose of this study is to confront the high precision combined HERA data \cite{Aaron:2009aa} with the b-CGC dipole model, in order to examine the effects of the tighter constraints on model parameters.   Since the IP-Sat dipole model was also recently updated with the recent combined HERA data \cite{ipsat-new}, we also compare the b-CGC and the IP-Sat results for both DIS and exclusive diffractive data at HERA, and provide predictions for various observable for a wide range of kinematics. Below, we summarize a few key results,  the details can be found in Ref.\,\cite{bcgc-new}.

\vspace{-0.2cm}
\section{Inclusive DIS and exclusive diffractive processes; a unified description}
\vspace{-0.2cm}
In the dipole picture, the scattering amplitude for the exclusive diffractive process $\gamma^*+p\to E+p$ with a final-state vector meson $E=J/\Psi, \phi,\rho$  or a real photon $E=\gamma$  in DVCS, can be written in terms of a convolution of the $q\bar{q}$ dipole-proton scattering amplitude $\mathcal{N}$ and the overlap wave-functions of photon and the exclusive final-state particle $\Psi_{E}^{*}\Psi$  \cite{bcgc-old,bcgc-new}, 
\begin{equation} \label{am-i}
  \mathcal{A}^{\gamma^* p\rightarrow Ep}_{T,L} (x,Q, \Delta)= \mathrm{2i}\,\int d^2\vec{r}\int_0^1 dz \int d^2\vec{b}\;(\Psi_{E}^{*}\Psi)_{T,L}\;\mathrm{e}^{-\mathrm{i}[\vec{b}-(1-z)\vec{r}]\cdot\vec{\Delta}}\mathcal{N}\left(x,r,b\right), 
\end{equation}
where $\vec{\Delta}^2=-t$ with $t$ being the squared momentum transfer, $r$ and b denote the dipole transverse size and impact-parameter of the collision, respectively. The differential cross-section for the exclusive diffractive process can be then given,
\begin{equation}
 \frac{d\sigma^{\gamma^* p\rightarrow Ep}_{T,L}}{d t}  = \frac{1}{16\pi}\left\lvert\mathcal{A}^{\gamma^* p\rightarrow Ep}_{T,L}\right\rvert^2\;(1+\beta^2)R_g^{2},
  \label{vm}
\end{equation}
with 
\bea \label{eq:beta} 
  \beta &=& \tan\left(\frac{\pi\delta}{2}\right), \hspace{0.5cm} R_g(\delta) = \frac{2^{2\delta+3}}{\sqrt{\pi}}\frac{\Gamma(\delta+5/2)}{\Gamma(\delta+4)}, \hspace{0.5cm}
\delta \equiv \frac{\partial\ln\left(\mathcal{A}_{T,L}^{\gamma^* p\rightarrow Vp}\right)}{\partial\ln(1/x)}. \
\eea 
The total deeply inelastic cross-section for a given $x$ and $Q^2$ can be obtained from \eq{am-i},
\begin{equation}\label{gp}
  \sigma_{L,T}^{\gamma^*p}(Q^2,x) =Im   \mathcal{A}^{\gamma^* p\rightarrow Ep}_{T,L} (x,Q, \Delta=0).
\end{equation}
The proton structure function $F_2$, the longitudinal structure function $F_L$  and reduced cross-section $\sigma_{r}$  can be then written in terms of the total $\gamma^{\star}p$ cross-section \cite{bcgc-new}. As seen in Eqs.\,(\ref{am-i}, \ref{vm}), $|t|$ and $b$ are directly related and the impact-parameter dependence of the dipole amplitude is crucial for describing  exclusive diffractive processes. A simple $b$-dependence for the dipole amplitude is obtained by combining the Glauber-Mueller form \cite{ipsat-old,ipsat-new}  of the amplitude
\bea
\mathcal{N}\left(x,r,b\right)  &=&1-\exp\left(-\frac{\pi^{2}r^{2}}{2N_{c}}\alpha_{s}\left(\mu^{2}\right)xg\left(x,\mu^{2}\right)T_{G}(b)\right), \label{ip-sat} 
\eea
where $xg\left(x,\mu^{2}\right)$ is the gluon density evolved up to  the scale $\mu$ with leading-order (LO) DGLAP gluon evolution. The scale $\mu^2$ is related to the dipole transverse size by
$\mu^{2}=4/r^{2}+\mu_{0}^{2} $ and the initial gluon distribution at the scale $\mu_0^2$,  and the impact parameter profile are taken to be,
\bea
xg\left(x,\mu_{0}^{2}\right) =A_{g}\,x^{-\lambda_{g}}(1-x)^{5.6},  \hspace{1cm} T_{G}(b)&=& \frac{1}{2\pi B_G}\exp\left(-b^2/2B_G\right). \label{g} 
\eea
 The parameter $B_G$ will be fixed with experimental data for exclusive $J/\Psi$ production. We take the corresponding one loop running-coupling value of $\alpha_s$ with $\Lambda_{\text{QCD}}=0.156$ GeV fixed by the experimentally measured value of $\alpha_s$ at the $Z^0$ mass. 
The parameters $A_{g},\lambda_{g}, \mu_{0}^{2}$ and $B_G$ are the only free parameters of our model which will be fixed by a fit to the reduced cross-section \cite{ipsat-new}. 

\begin{figure}[t]       
\includegraphics[width=0.48\textwidth,clip]{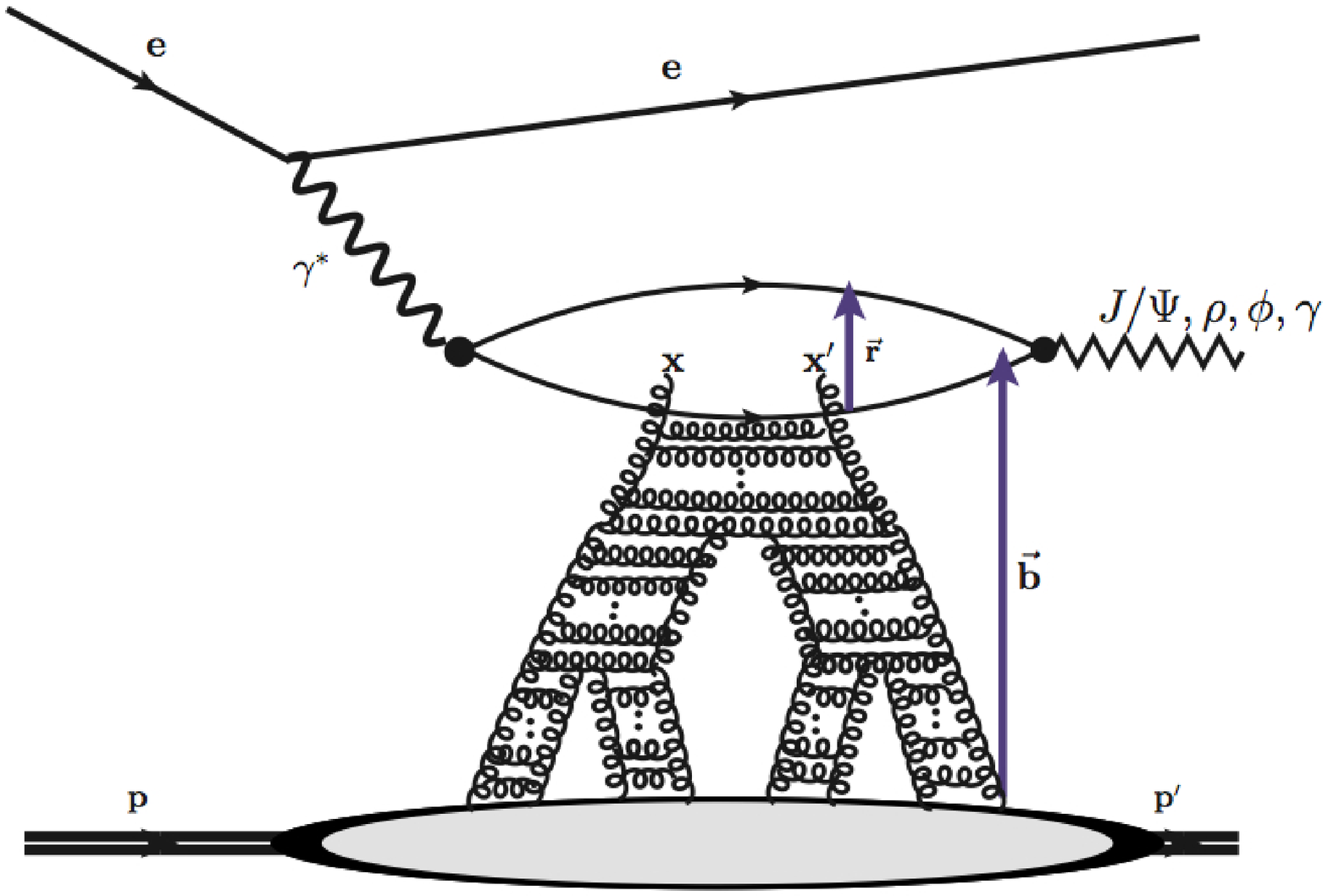}   
\includegraphics[width=0.48\textwidth,clip]{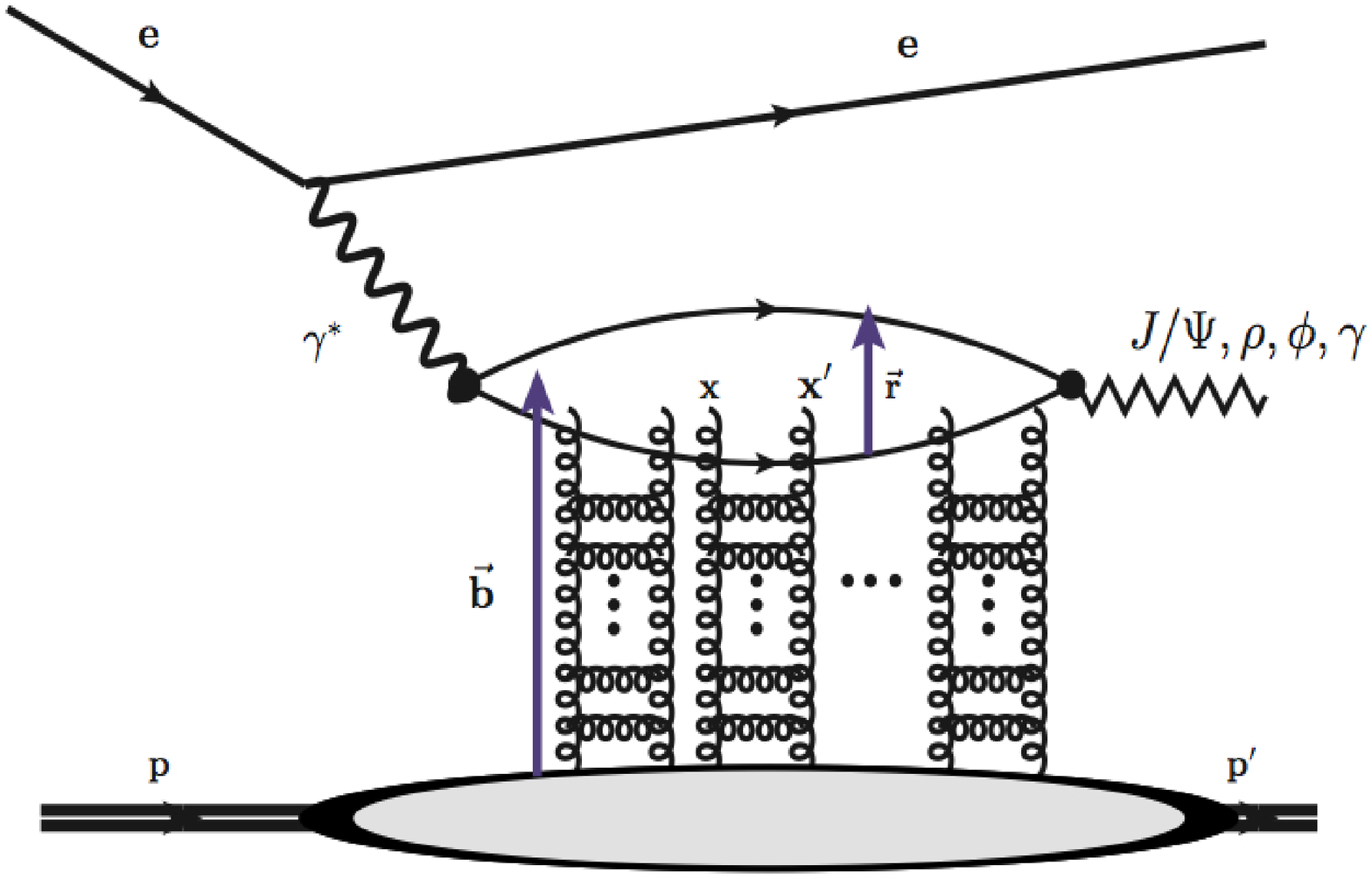}   
 \caption{The exclusive diffractive processes (with  $p\ne p^{\prime}$ or $t\ne 0$, and $x<<x^{\prime}<<1$)  in the b-CGC dipole model (left) and the IP-Sat dipole model (right) in the rest frame of the target.  }
\label{f-f}           
\end{figure}   

\begin{figure}[t]       
\includegraphics[width=0.37\textwidth,clip]{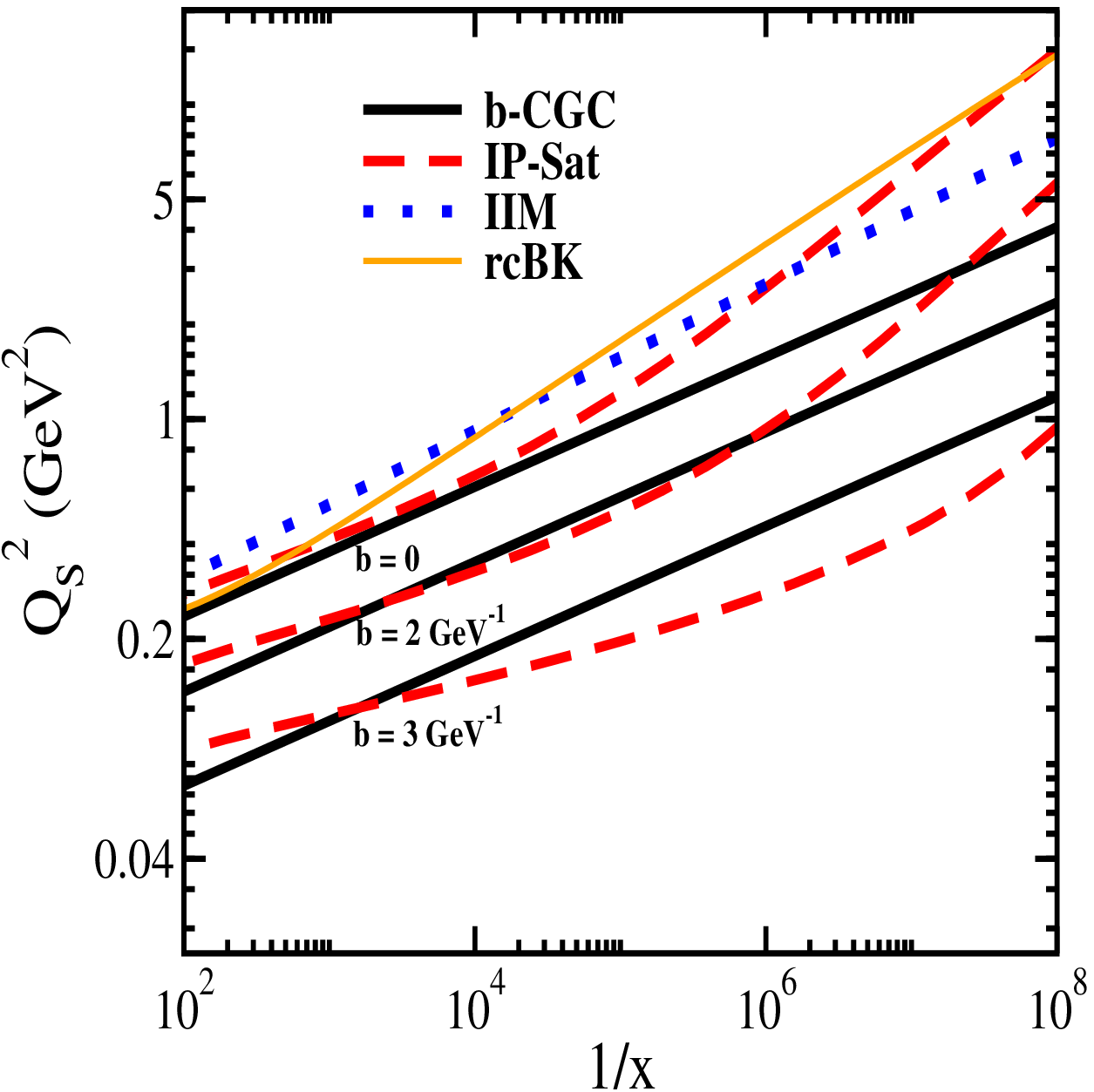}    
\includegraphics[width=0.43\textwidth, height=0.25 \textheight, clip]{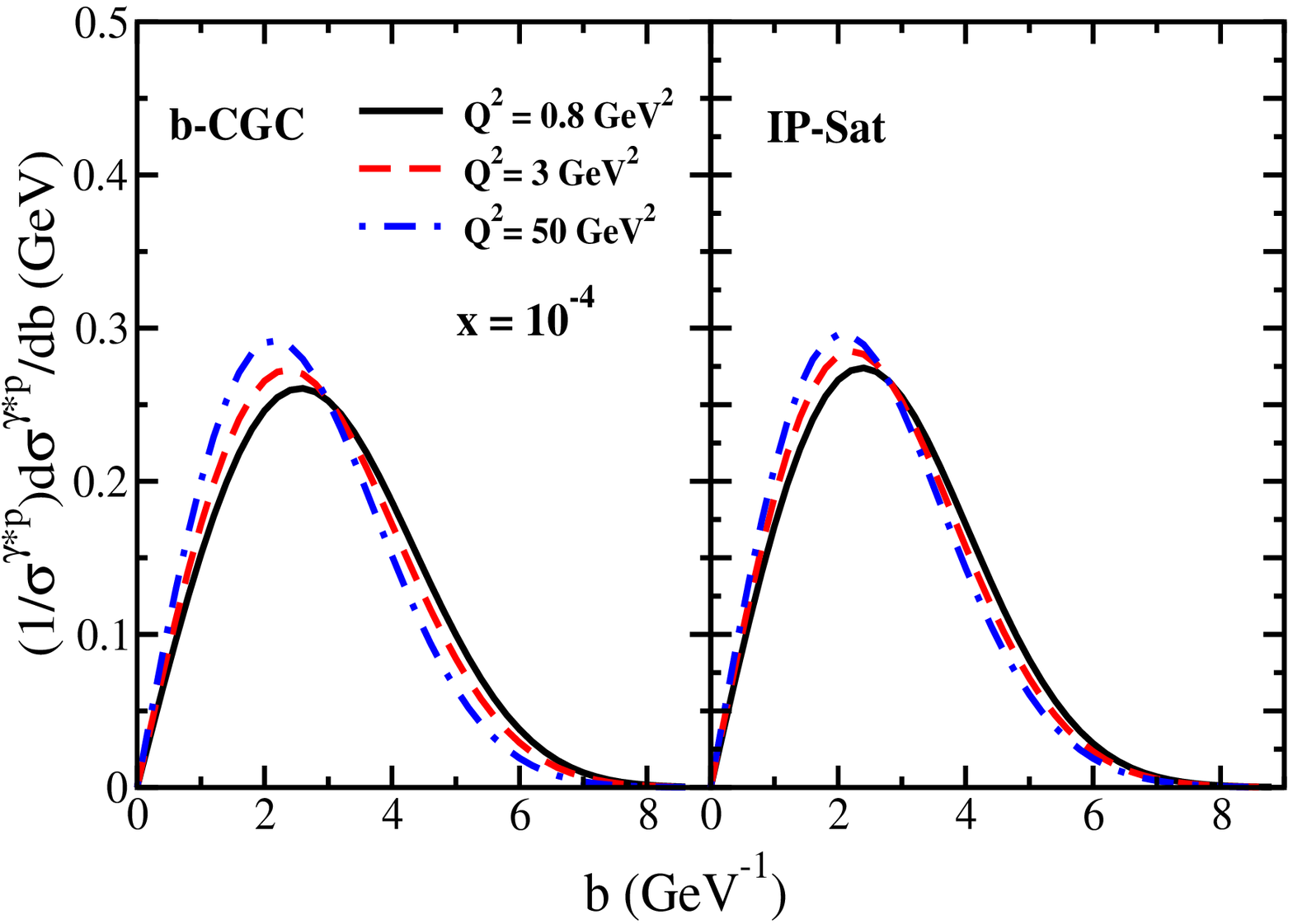}   
\caption{Left: The saturation scale $Q_S$ in the b-CGC and the IP-Sat dipole models, as a function of $1/x$, at various impact-parameters $b$. For comparison we also show the impact-parameter independent saturation scale obtained from the Iancu-Itakura-Munier (IIM) model \cite{bcgc-new,IIM} and the running-coupling Balitsky-Kovchegov (rcBK) equation  \cite{rcbk}.  Right: The impact-parameter $b$ dependence of the total $\gamma^{\star}p$ cross-section, at  fixed $x$ and various $Q^2$, in the b-CGC and the IP-Sat dipole models. }
\label{f-g2}           
\end{figure}    

The b-CGC model \cite{bcgc-old,bcgc-new} is constructed by smoothly interpolating between two limiting behaviors which are analytically under control, namely the solution to the BFKL equation in the vicinity of the saturation line for small dipole sizes, $r <<1/Q_s$, and the Levin-Tuchin solution \cite{LT} of the Balitsky-Kovchegov (BK) equation \cite{bk} deep inside the saturation region for larger dipoles, $r >>1/Q_s$ \cite{IIM}. In the b-CGC dipole model, the color dipole-proton amplitude is given by \cite{IIM}, 
\bea \label{CA5}
N\Lb x, r, b\Rb\,\,=\,\, \left\{\begin{array}{l}\,\,\,N_0\,\Lb \frac{r Q_s}{2}\Rb^{2\gamma_{eff}}\,\,\,\,\,\,\,\,r Q_s\,\leq\,2\,,\\ \\
1\,\,-\,\,\exp\Lb -\mathcal{A} \ln^2\Lb \mathcal{B} r Q_s\Rb\Rb\,\,\,\,\,\,\,\,\,\,\,\,\,\,\,\ rQ_s\,>\,2\,,\end{array}
\right.
\eea
with impact-parameter dependent effective anomalous dimension and the scale $Q_s$ \cite{bcgc-old} defined as  
\beq \label{g-eff}
\gamma_{eff}=\gamma_s\,\,+\,\,\frac{1}{\kappa \lambda Y}\ln\Lb\frac{2}{r Q_s}\Rb, \hspace{0.5cm} Q_s=\,\,\Lb \frac{x_0}{x}\Rb^{\frac{\lambda}{2}}\,\exp\left\{- \frac{b^2}{4\gamma_s B_{CGC}}\right\} \text{GeV},
\end{equation}
where  $Y=\ln(1/x)$ and $\kappa= \chi''(\gamma_s)/\chi'(\gamma_s)$, with $\chi$ being the LO BFKL
characteristic function. The second term (diffusion term)  in $\gamma_{eff}$ enhances the anomalous dimension from its value at BFKL $\gamma_{eff}\to\gamma_s$ to DGLAP  $\gamma_{eff}\to  1$, matching the BFKL region to the color-transparency regime of the DGLAP for small dipole sizes\footnote{Notice that the anomalous dimension defined via \eq{g-eff} is not well-defined as $r\to 0$. However, this limiting case has negligible contribution to the total cross-section.} (or high virtualities).
The parameters $\mathcal{A}$ and $\mathcal{B}$ in \eq{CA5} are determined uniquely from the matching of the dipole amplitude and its logarithmic derivatives at $rQ_s=2$. In the b-CGC dipole model we let the parameter $N_0$ to be free along with $\gamma_s, x_0, \lambda$, and obtain their values via a fit to the recent HERA combined data for the reduced cross-section \cite{bcgc-new}. Although both models include saturation effects and depend on impact-parameter, the former is based on the non-linear BK equation, while the latter is based on DGLAP evolution, incorporating the saturation effect via Glauber-Mueller approximation \cite{ipsat-old,ipsat-new}. Therefore, the underlying dynamics of two models are quite distinct. The difference between the b-CGC and the IP-Sat models  is illustrated in \fig{f-f}. 

 The extracted values of $\gamma_s\approx 0.65$ and $\lambda \approx 0.20$ from the new combined HERA \cite{bcgc-new} are now approximately compatible with the perturbative expectation, in drastic contrast to old fit in Ref.\,\cite{bcgc-old}.  Other key features of our novel fit \cite{bcgc-new} is that the preferred value of  light quark masses is close to the current quark masses $m_u\approx 10^{-2}\div 10^{-4}$, and also smaller value for the parameter $B_{CGC}$ in the impact-parameter profile of the saturation scale, compared to the old analysis.  In the IP-Sat model, the key features of new fit \cite{ipsat-new} include the preferred lower values for the light quark masses $m_u\approx 10^{-2}\div 10^{-4}$ and also positive value for the parameter $\lambda_g>0$ in \eq{g} which are in sharp contrast with the old fit in Ref.\,\cite{ipsat-old}.

\begin{figure}
  \includegraphics[width=0.33 \textwidth,clip]{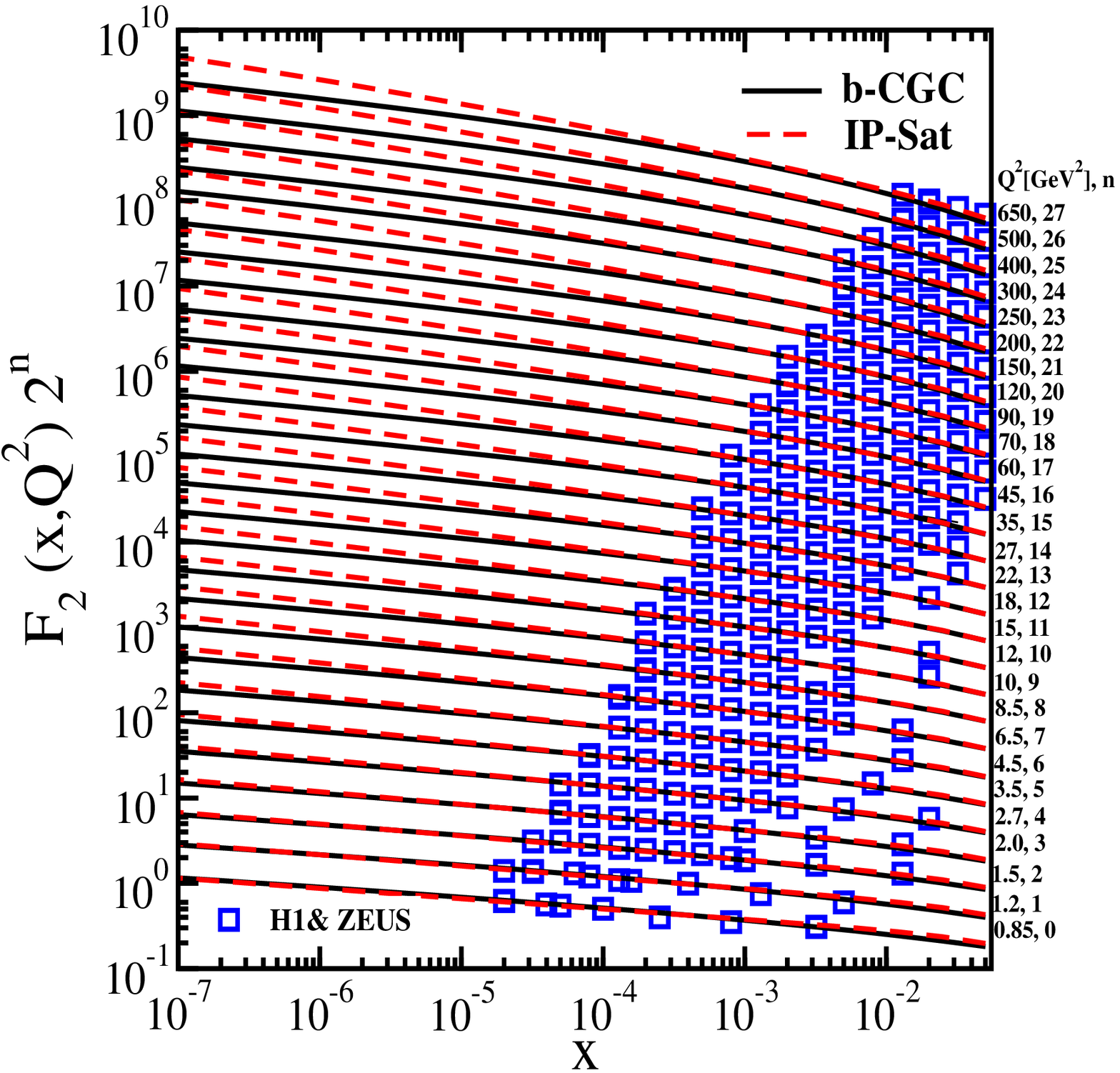}
\includegraphics[width=0.33\textwidth,clip]{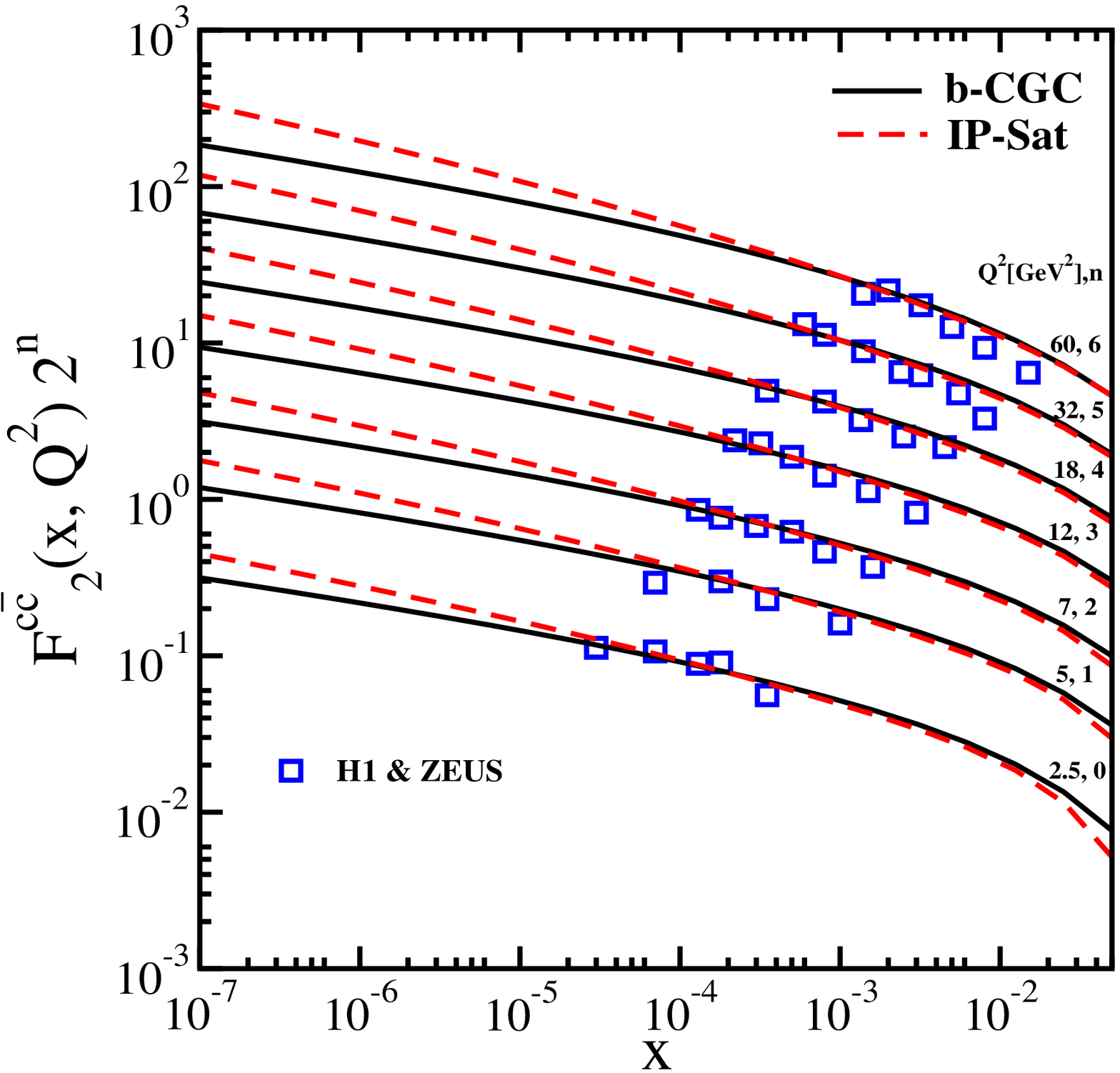}
\includegraphics[width=0.33\textwidth,clip]{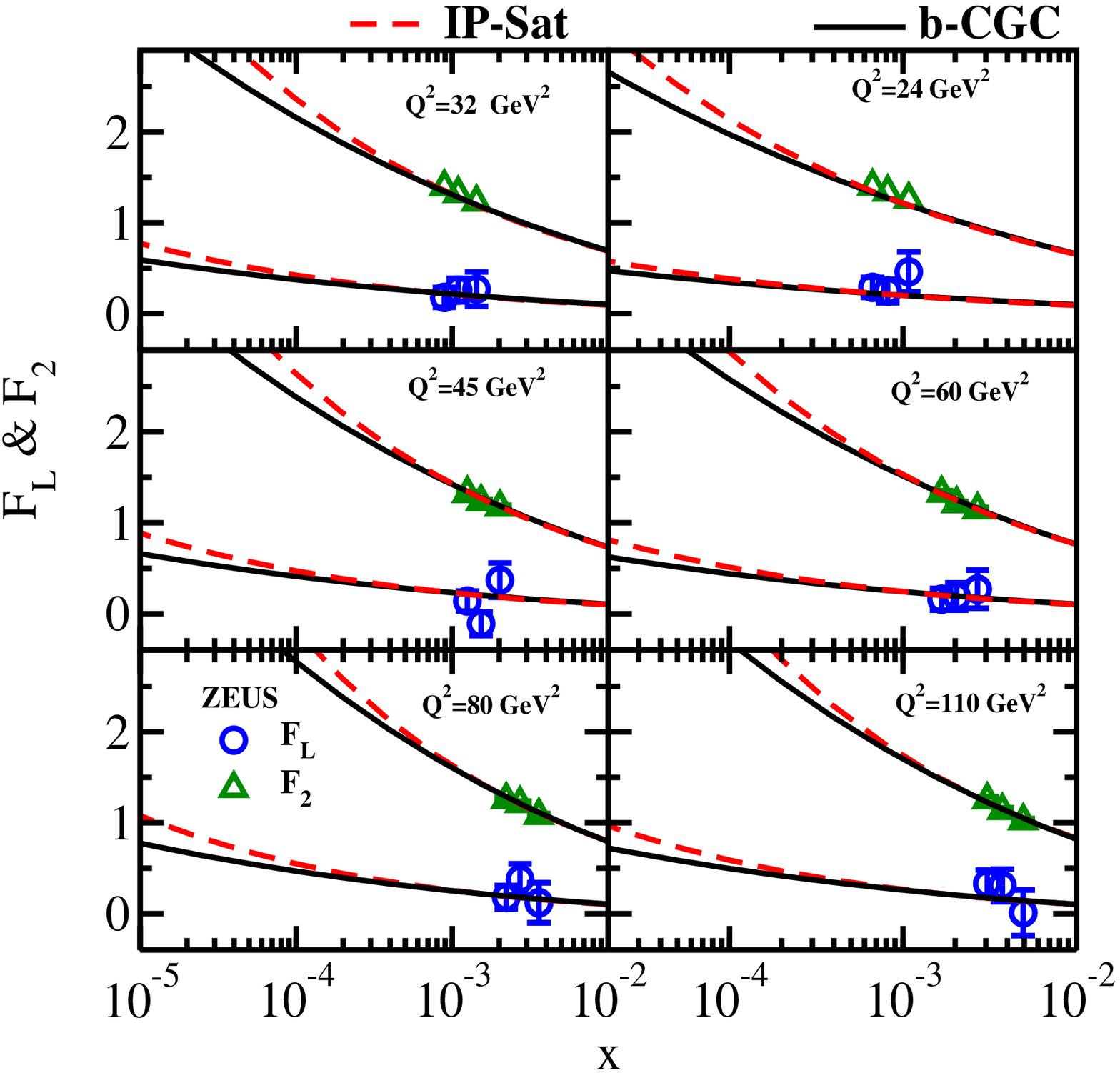}
  \caption{Results for the structure functions $F_2(x,Q^2)$, $F_2^{c\bar{c}}(x,Q^2)$ and $F_L(x,Q^2)$ as function of $x$, for various values of $Q^2$, in the b-CGC (solid line) and the IP-Sat (dashed line) dipole models. The plots are taken from Ref.\,\cite{bcgc-new}.}
  \label{f-f2}
\end{figure}

In \fig{f-g2} left panel, we show the impact-parameter $b$ dependence of the total $\gamma^{*}p$ cross-section calculated by the b-CGC and the IP-Sat dipole models, at fixed x and various $Q^2$.  We see that in both models the main contribution of the integrand in the structure functions and the reduced cross-section  at various virtualities $Q^2$ comes from $1\leq b [\text{GeV}^{-1}]\leq 4$. Although the $b$ dependence of the dipole amplitude is different in the b-CGC and the IP-Sat models, remarkably both lead to the same conclusion that the typical $b$ probed in the total $\gamma^{*}p$ (and the structure functions) is about  $2\div 3~\text{GeV}^{-1}$. We define the saturation scale $Q_S^2=2/r_S^2$, where $r_S$ is the saturation radius, as a scale where the dipole scattering amplitude has the value $\mathcal{N}(x,r_S=\sqrt{2}/Q_S,b)=1-\exp(-1/2)=0.4$ \cite{bcgc-old,bcgc-new}. Note that the saturation scale does not have a unique definition,  nevertheless, the above definition gives a useful baseline to compare relative magnitude of saturation scale in different models. In \fig{f-g2} (left), we show the saturation scale as a function of impact parameter $b$, for different values of $x$ in different models. We see that the saturation scale as a function of  $1/x$ grows relatively  faster for more central collisions ($b\approx 0$). Moreover, the saturation scale at different impact parameters can be significantly different, even by one order of magnitude. This non-trivial behavior shows the importance of the impact-parameter dependence of the saturation scale. It is remarkable that although the b-CGC and the IP-Sat models are different, both give similar saturation scales within the x-region that they have been fitted to the HERA data, namely within 
$x\in [10^{-2},10^{-5}]$. However, at smaller $x$ about $x<10^{-5}$, they become significantly different and that leads to sizeable different predictions for the structure functions (and other observables) at very small x as shown in \fig{f-f2}. 

With the parameters of the b-CGC and the IP-Sat model extracted from the $\chi$-squared fit to the reduced inclusive DIS cross-section, we then compute the structure functions $F_2(x,Q^2)$, the charm structure function  $F_2^{c\bar{c}}(x,Q^2)$, the longitudinal structure function $F_L(x,Q^2)$ and compare to the combined HERA data in \fig{f-f2}. Note that experimental data for $F_2$, $F_L$ and $F_2^{c\bar{c}}$  were not included in our fit and therefore this can be considered as a non-trivial consistency check of the model.

\begin{figure}[t]       
\includegraphics[width=0.3\textwidth,clip]{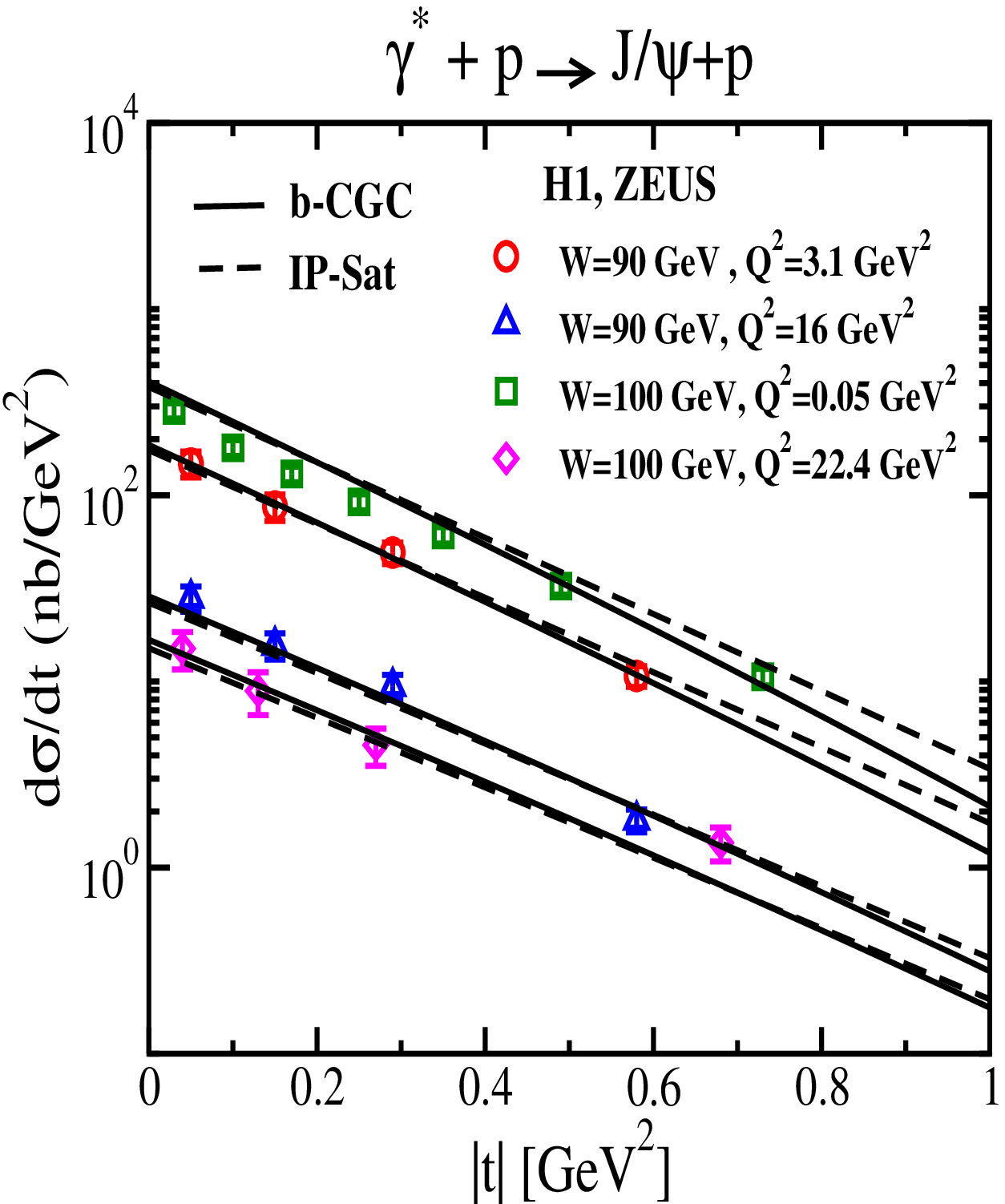}   
\includegraphics[width=0.3\textwidth,clip]{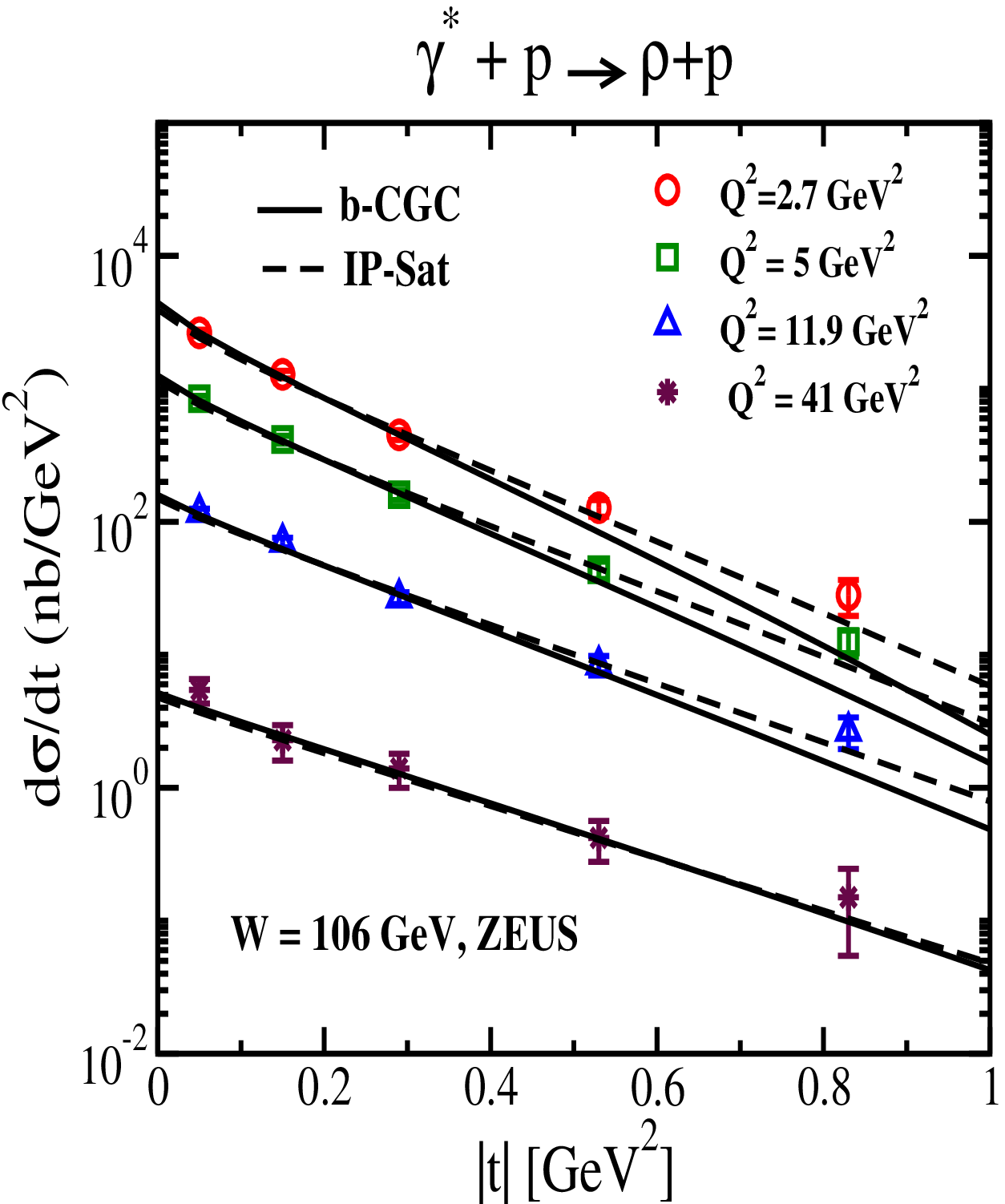}
\includegraphics[width=0.3\textwidth,clip]{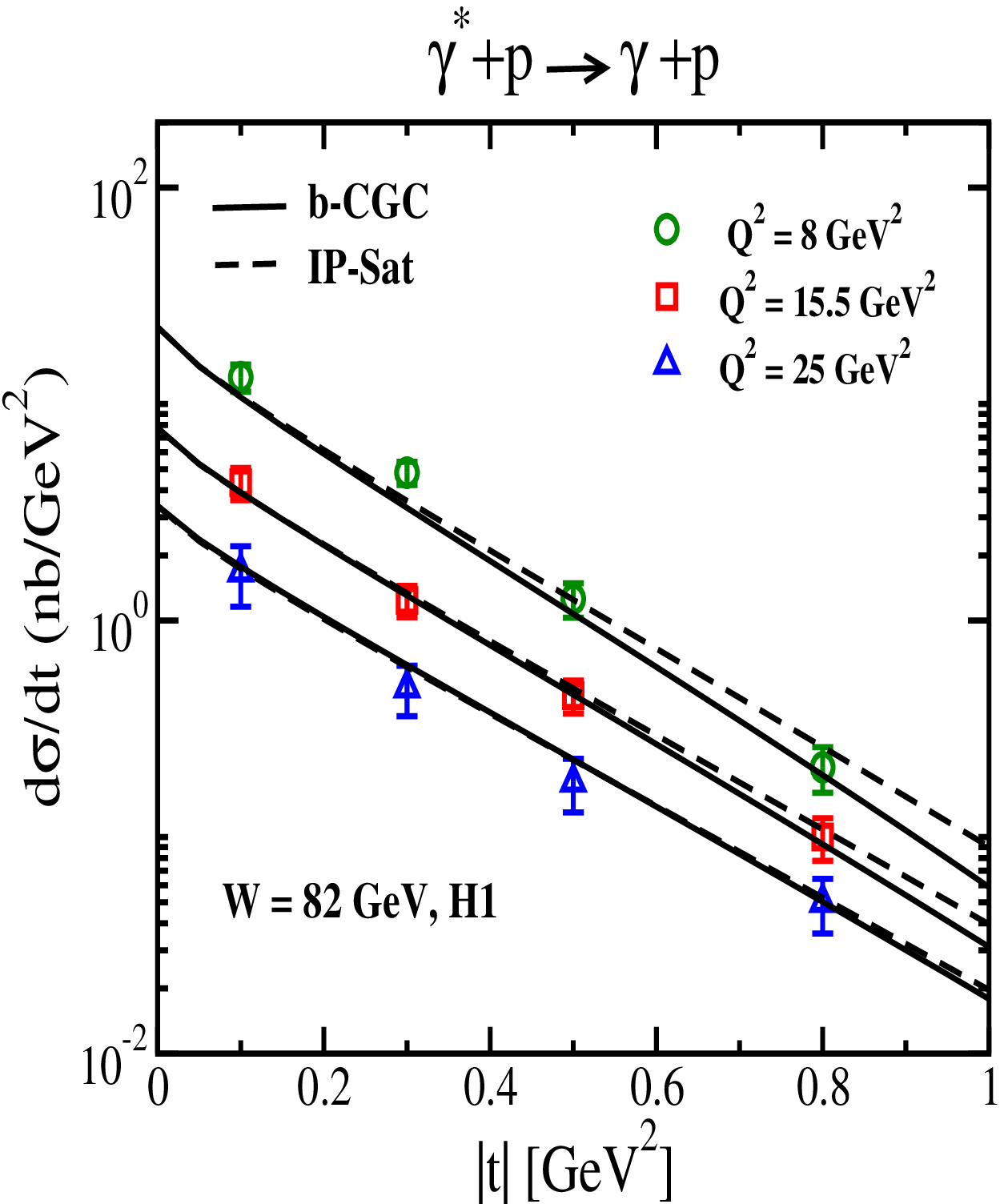}
 \caption{Differential vector meson cross-sections for $J/\Psi$, $\rho$ and DVCS, as a function of $|t|$. Data for a given $W$ with varying $Q^2$, are compared to the results from the b-CGC (solid lines) and IP-Sat (dashed lines) models, using the parameter sets with $m_c=1.27$ GeV in both models. The plots are taken from Ref.\,\cite{bcgc-new}.  }
  \label{f-vt}
\end{figure}

In \fig{f-vt}, we show the $t$-distribution of  exclusive vector mesons production and DVCS obtained by using the b-CGC and IP-Sat models. We fix the width of the impact-parameter profile of the saturation scale via a fit to the slope of the $t$-distribution of the diffractive $J/\Psi$ production at low $t$ (at a fixed $W$ and $Q^2$) , the other data points shown in \fig{f-vt} were not included into the fit. It is seen that the model predictions for the $t$ distribution becomes different at large $t$ where we do not have currently data. Note that large $t$ corresponds to small $b$. On the other hand, as we already stressed the typical $b$ probes in DIS is not central, see \fig{f-g2}, as a result the saturation models are less constrained at large $|t|$.      

For comparison of our results with other observables at HERA and the LHC, see Refs.\,\cite{bcgc-new,Nestor-VM}. It is remarkable that with only 4 parameters fixed to reduced cross-section, the b-CGC and IP-Sat models give excellent description of almost all available data on inclusive and exclusive diffractive processes at HERA at small-x ($x\le 10^{-2}$). The b-CGC and also IP-Sat models have been intensively applied to various reactions including heavy ion collisions. However, the parameters employed in these studies were determined from data from H1 and ZEUS predating the combined data sets for the proton. It remains to be seen what the impact of the new fits are on final state observables in heavy ion collisions. For example, it has been recently shown that while the old (2008) b-CGC fit, does not provide a good description of the diffractive photoproduction data at the LHC, the new b-CGC fit remarkably agrees with the recent LHC data \cite{Nestor-VM,na-VM,magno-VM}.

\end{document}